\documentclass[
reprint,
superscriptaddress,
bibnotes,
amsmath,amssymb,
aps,
pra,
]{revtex4-1}

\usepackage{graphicx,import}
\usepackage{dcolumn}
\usepackage{braket}
\usepackage{textcomp, gensymb}
\usepackage{bm}
\usepackage{amsmath}
\usepackage{float}
\usepackage{mathtools}
\usepackage{color}
\usepackage[dvipsnames,table,xcdraw]{xcolor}
\usepackage[utf8]{inputenc}
\usepackage{csquotes}
\usepackage{thmtools}

\usepackage{dsfont}
\usepackage{bbm}

\usepackage{hyperref}

\usepackage[normalem]{ulem}

\usepackage{outline}
\usepackage{physics}
\usepackage{multirow}
\usepackage{siunitx}
\usepackage{tikz}
\usetikzlibrary{shapes.geometric, arrows,positioning}
\usepackage{qcircuit}
\usepackage{lipsum}
\usepackage{chemfig}
\usepackage{silence}
\usepackage{algorithm}
\usepackage{algpseudocode}

\makeatletter
\def\p@subsection{}
\makeatother
\makeatletter
\def\p@subsubsection{}
\makeatother

\allowdisplaybreaks


\begin{document}

\title{Concatenated continuous-drive protection of exchange-based spin-qubit gates}

\author{Takuma Kuno}
\email{takuma.kuno.pg@hitachi.com}
\affiliation{Research and Development Group, Hitachi, Ltd., Kokubunji, Tokyo 185-8601, Japan}
\affiliation{Department of Electrical and Electronic Engineering, Institute of Science Tokyo, Meguro, Tokyo
    152-8552, Japan}

\author{Takeru Utsugi}
\affiliation{Research and Development Group, Hitachi, Ltd., Kokubunji, Tokyo 185-8601, Japan}

\author{Andrew J. Ramsay}
\affiliation{Hitachi Cambridge Laboratory, J. J. Thomson Ave., Cambridge, CB3
    0HE, United Kingdom}

\author{Hideo Arimoto}
\affiliation{Research and Development Group, Hitachi, Ltd., Kokubunji, Tokyo 185-8601, Japan}

\author{Tetsuo Kodera}
\affiliation{Department of Electrical and Electronic Engineering, Institute of Science Tokyo, Meguro, Tokyo
    152-8552, Japan}

\author{Hiroyuki Mizuno}
\affiliation{Research and Development Group, Hitachi, Ltd., Kokubunji, Tokyo 185-8601, Japan}

\date{\today}

\begin{abstract}

    Exchange-based two-qubit gates in semiconductor spin qubits are
    sensitive to low-frequency detuning noise and control errors.
    Here, we extend concatenated continuous driving (CCD), implemented by
    phase modulation of a global microwave drive, to exchange-based SWAP and
    $\sqrt{\mathrm{SWAP}}$ operations between
    double-dressed spin qubits and investigate its effectiveness for
    protecting two-qubit gates. We find that CCD retains strong
    protection against quasistatic common-mode detuning errors while substantially
    improving robustness against Rabi-frequency errors and qubit-to-qubit
    Rabi-frequency mismatch compared with conventional continuous dressing.
    By contrast, the gate fidelity is impacted by differential detuning errors.
    By modeling partially correlated quasistatic detuning noise, we
    further show that the noise-averaged gate fidelity depends strongly on
    the spatial correlation of the qubit-frequency fluctuations, with
    positive correlations enhancing CCD protection by suppressing the
    differential-noise component. These results identify the symmetry and
    spatial correlation of qubit-frequency noise as key factors governing
    continuous-drive protection in coupled spin-qubit systems.

\end{abstract}

\maketitle
\section{Introduction}
Achieving robust, high-fidelity two-qubit control is a central
challenge for scalable silicon spin-qubit processors~\cite{huang2019fidelity,xue2022quantum,mills2022two,noiri2022fast}.
A major limitation arises from fluctuations that perturb both the
qubit resonance frequencies and the exchange interaction.
Isotopic enrichment strongly suppresses hyperfine noise in silicon~\cite{witzel2010electron,itoh2014isotope}, although residual
$^{29}\mathrm{Si}$ nuclear spins have been observed to contribute to
qubit-frequency fluctuations in enriched $^{28}\mathrm{Si}$
devices~\cite{zhao2019single}.
Electrical noise can shift the qubit resonance frequency through
spin--electric coupling~\cite{yoneda2018quantum,park2025passive,capannelli2025tracking}.
Because the exchange interaction is electrically controlled, charge
noise can also modulate the exchange coupling between neighboring
spins~\cite{reed2016reduced,connors2022charge}.
These noise channels can therefore degrade exchange-based two-qubit
gates through both single-qubit frequency fluctuations and fluctuations
of the exchange interaction~\cite{throckmorton2017environmental,shehata2023modeling}.
The isotropic Heisenberg exchange interaction provides a natural
mechanism for implementing SWAP and entangling
$\sqrt{\mathrm{SWAP}}$ operations between neighboring spin qubits,
as established in foundational proposals for quantum-dot spin
qubits~\cite{loss1998quantum,burkard2023semiconductor}.
Coherent SWAP operations have subsequently been demonstrated in
silicon quantum-dot devices~\cite{sigillito2019coherent,ni2025swap}.
Suppressing noise and control errors during such exchange-based
operations is therefore important for realizing high-fidelity
two-qubit control.

Continuous driving provides an effective approach to suppressing
low-frequency qubit-frequency noise. Under a resonant microwave (MW)
field, spin qubits form dressed states whose energy splitting is
primarily determined by the Rabi frequency, thereby reducing their
sensitivity to quasistatic longitudinal noise~\cite{laucht2017dressed}.

Conventional dressing, however, remains sensitive to fluctuations
of the Rabi frequency itself. Concatenated continuous driving (CCD)
addresses this limitation by introducing an additional protection
field in the dressed-state basis~\cite{cai2012robust,ramsay2023coherence}.
In particular, phase modulation provides a practical means of generating
the secondary continuous-driving field~\cite{farfurnik2017experimental}.
In silicon spin qubits, phase-modulated CCD (PMCCD) has been shown to
produce a double-dressed-state basis and provide protection against both
detuning and Rabi-frequency errors~\cite{kuno2026robust,kuno2026concatenated}.

Global-control architectures based on continuously dressed spin qubits
have been proposed as a scalable approach to single- and two-qubit
control~\cite{seedhouse2021quantum}.
Exchange-based two-qubit operations under continuous global driving
have subsequently been demonstrated in silicon spin qubits, showing that
exchange dynamics can be maintained while the global drive remains
applied~\cite{hansen2024entangling}.
Under concatenated continuous driving, however, the exchange interaction
acts between double-dressed qubits in the second rotating frame, where
the secondary drive introduces an additional protection energy gap.
A central question is therefore whether this additional protection
remains effective while the exchange interaction simultaneously
generates an entangling operation. For two-qubit gates, the spatial
structure of the noise is also important, because fluctuations
experienced by neighboring qubits can be spatially correlated~\cite{yoneda2023noise,rojas2023spatial}.
Such correlations influence the relative weight of common-mode and
differential fluctuations, which can affect the relative dynamics of
exchange-coupled spins differently.

Here, we theoretically and numerically investigate
SWAP and $\sqrt{\mathrm{SWAP}}$ operations between double-dressed spin
qubits protected by PMCCD. We consider a common phase-modulated MW field
while allowing qubit-dependent detuning and Rabi-frequency errors, as
relevant to globally driven spin-qubit architectures.
We find that PMCCD retains strong protection against common-mode
detuning and substantially improves robustness against Rabi-frequency
errors and qubit-to-qubit Rabi-frequency mismatch compared with
conventional continuous dressing. In contrast, differential detuning
remains a significant source of gate error. By extending the analysis
to partially correlated quasistatic detuning noise, we further show
that the noise-averaged gate fidelity depends strongly on the spatial
correlation of the qubit-frequency fluctuations: positive correlations
reduce the differential-noise component and substantially enhance the
effectiveness of PMCCD protection. These results identify the symmetry
and spatial correlation of qubit-frequency noise as key factors
governing continuous-drive protection of exchange-based entangling
gates.

\section{Exchange-based two-qubit operations with PMCCD-protected qubits}
\label{sec:model}

\begin{figure}[hbt]
    \includegraphics[width=\columnwidth]{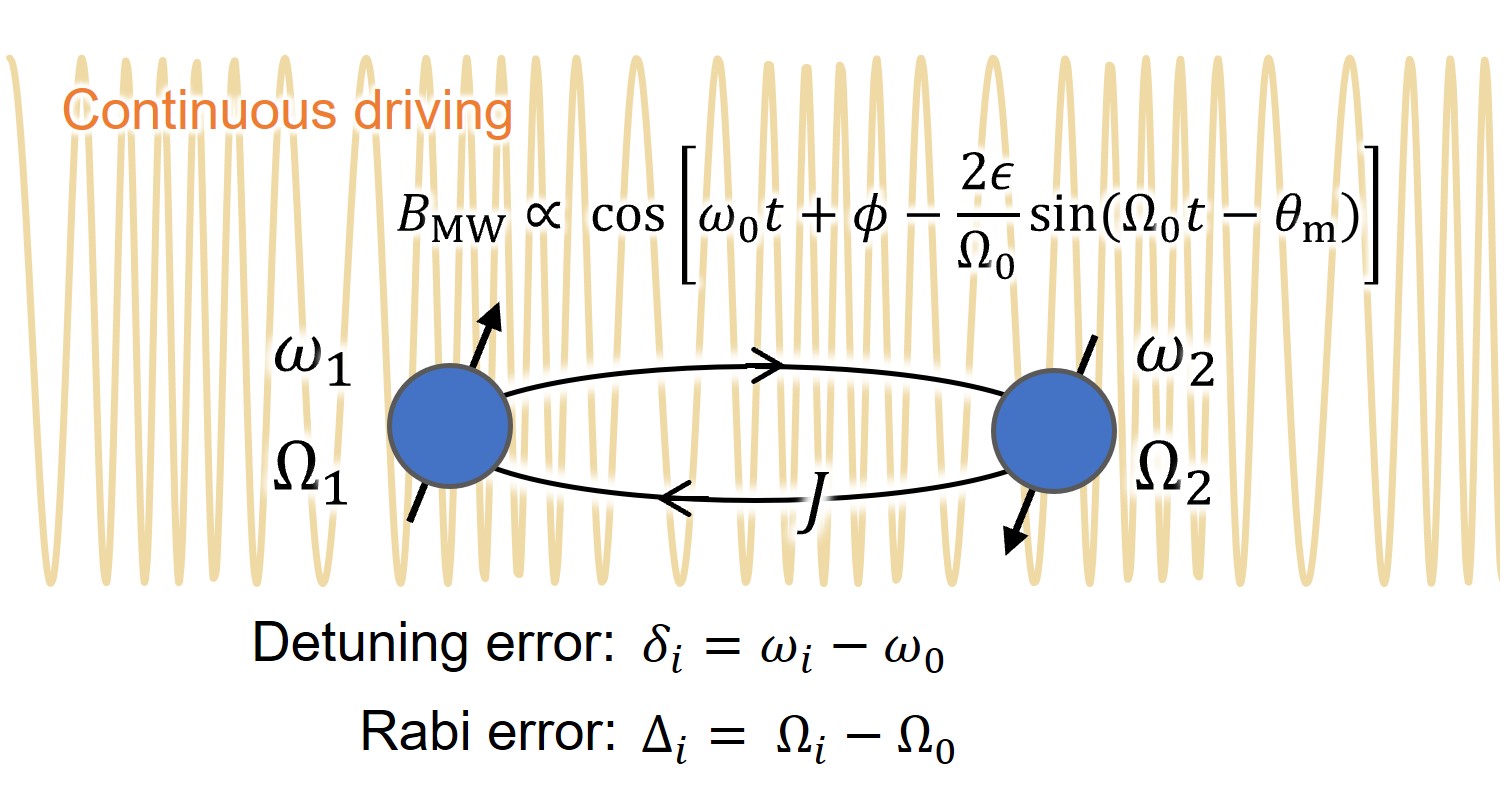}
    \caption{
        {\bf Schematic of a SWAP operation between PMCCD-protected
                double-dressed spin qubits.}
        A global phase-modulated MW field [Eq.~\eqref{eq:MW_waveform}]
        creates the dressed and double-dressed qubit states, while the
        exchange interaction generates a SWAP-type two-qubit operation.
    }
    \label{Concept}
\end{figure}

We consider two exchange-coupled spin qubits subject to a global
phase-modulated MW field (Fig.~\ref{Concept}). The same MW field is applied to
both qubits, while their effective Rabi angular frequencies,
$\Omega_1$ and $\Omega_2$, may differ owing to spatial variations of
the MW field or qubit-dependent coupling to the global
drive~\cite{hansen2021pulse,vahapoglu2022coherent}.
The MW field is applied with a sinusoidally modulated phase,
\begin{equation}
    B_{\mathrm{MW}}(t)
    \propto
    \cos\!\left[
        \omega_0 t + \phi
        - \frac{2\epsilon}{\Omega_0}
        \sin(\Omega_0 t-\theta_{\mathrm{m}})
        \right],
    \label{eq:MW_waveform}
\end{equation}
where $\omega_0$ is the carrier angular frequency and $\phi$ is the
initial MW phase. The phase is modulated at the nominal Rabi angular
frequency $\Omega_0$ with modulation strength $\epsilon$ and modulation
phase $\theta_{\mathrm{m}}$.
Throughout this work, the two-qubit interaction is modeled by an
isotropic Heisenberg exchange coupling.
Setting $\hbar=1$, the laboratory-frame Hamiltonian is
\begin{align}
    H_{\mathrm{lab}}(t)
     & =
    \frac{\omega_1}{2}\,\sigma_z \otimes I
    +
    \frac{\omega_2}{2}\,I \otimes \sigma_z
    \nonumber \\
     & \quad
    + \Omega_1
    \cos\!\left[
        \omega_0 t + \phi
        - \frac{2\epsilon}{\Omega_0}
        \sin(\Omega_0 t-\theta_{\mathrm{m}})
        \right]
    \sigma_x \otimes I
    \nonumber \\
     & \quad
    + \Omega_2
    \cos\!\left[
        \omega_0 t + \phi
        - \frac{2\epsilon}{\Omega_0}
        \sin(\Omega_0 t-\theta_{\mathrm{m}})
        \right]
    I \otimes \sigma_x
    \nonumber \\
     & \quad
    + \frac{J}{4}
    \sum_{\alpha=x,y,z}
    \sigma_\alpha \otimes \sigma_\alpha .
    \label{eq:Hlab_twoqubit_PMCCD}
\end{align}
Here, $\omega_1$ and $\omega_2$ are the Larmor angular frequencies of the two qubits and $J$ is the isotropic exchange coupling.
The case $\epsilon=0$ corresponds to conventional continuous dressing, whereas a finite $\epsilon$ introduces the secondary driving field required for PMCCD~\cite{kuno2026robust,kuno2026concatenated}.
Alternative implementations of the secondary field by amplitude or combined amplitude--phase modulation are compared in Appendix~B.

We first transform to a rotating frame that follows the phase-modulated MW field.
The corresponding unitary transformation is
\begin{equation}
    U_1(t)
    =
    \exp\!\left[
        -\frac{i}{2}
        \Phi_{\mathrm{CCD}}(t)
        \left(
        \sigma_z \otimes I + I \otimes \sigma_z
        \right)
        \right],
    \label{eq:U1_PMCCD_twoqubit}
\end{equation}
where
\begin{equation}
    \Phi_{\mathrm{CCD}}(t)
    =
    \omega_0 t
    -
    \frac{2\epsilon}{\Omega_0}
    \sin(\Omega_0 t-\theta_{\mathrm{m}}).
    \label{eq:Phi_CCD_twoqubit}
\end{equation}
The Hamiltonian in the first rotating frame is
\begin{equation}
    H_{\mathrm{rot}}^{(1)}(t)
    =
    U_1^\dagger(t)H_{\mathrm{lab}}(t)U_1(t)
    -
    iU_1^\dagger(t)\dot{U}_1(t).
    \label{eq:Hrot1_def_twoqubit}
\end{equation}
Applying the rotating-wave approximation (RWA) with respect to the carrier frequency, under the condition
\begin{equation}
    \omega_0
    \gg
    \Omega_i,\Omega_0,\epsilon,|\delta_i|,J,
\end{equation}
gives
\begin{align}
    H_{\mathrm{rot}}^{(1)}(t)
     & =
    \frac{\delta_1}{2}\,\sigma_z \otimes I
    +
    \frac{\delta_2}{2}\,I \otimes \sigma_z
    \nonumber \\
     & \quad
    + \frac{\Omega_1}{2}\,\sigma_\phi \otimes I
    +
    \frac{\Omega_2}{2}\,I \otimes \sigma_\phi
    \nonumber \\
     & \quad
    + \epsilon\cos(\Omega_0 t-\theta_{\mathrm{m}})
    \left(
    \sigma_z \otimes I + I \otimes \sigma_z
    \right)
    \nonumber \\
     & \quad
    + \frac{J}{4}
    \sum_{\alpha=x,y,z}
    \sigma_\alpha \otimes \sigma_\alpha ,
    \label{eq:Hrot1_twoqubit_PMCCD}
\end{align}
where
\begin{equation}
    \delta_i = \omega_i-\omega_0
    \label{eq:detuning_definition}
\end{equation}
is the detuning of qubit $i$ from the MW carrier frequency, and
\begin{equation}
    \sigma_\phi
    =
    \cos\phi\,\sigma_x
    +
    \sin\phi\,\sigma_y .
\end{equation}
Equation~\eqref{eq:Hrot1_twoqubit_PMCCD} shows that the phase modulation is converted into an oscillating field along the $\sigma_z$ direction in the first rotating frame.
This field constitutes the secondary drive of the concatenated continuous-driving protocol.

We next transform to a second rotating frame at the nominal Rabi
angular frequency $\Omega_0$,
\begin{equation}
    U_2(t)
    =
    \exp\!\left[
        -\frac{i}{2}
        \Omega_0 t
        \left(
        \sigma_\phi \otimes I
        +
        I \otimes \sigma_\phi
        \right)
        \right].
    \label{eq:U2_PMCCD_twoqubit}
\end{equation}
The Hamiltonian in this frame is
\begin{equation}
    H_{\mathrm{rot}}^{(2)}(t)
    =
    U_2^\dagger(t)H_{\mathrm{rot}}^{(1)}(t)U_2(t)
    -
    iU_2^\dagger(t)\dot{U}_2(t).
    \label{eq:Hrot2_def_twoqubit}
\end{equation}
We define the corresponding single-qubit rotation
$u_2(t)=\exp[-i\Omega_0 t\,\sigma_\phi/2]$, such that
$U_2(t)=u_2(t)\otimes u_2(t)$, and introduce
\begin{equation}
    \sigma_{\phi+\pi/2}
    =
    -\sin\phi\,\sigma_x
    +
    \cos\phi\,\sigma_y .
\end{equation}
We then use
\begin{align}
    u_2^\dagger(t)\sigma_z u_2(t)
     & =
    \cos(\Omega_0 t)\,\sigma_z
    +
    \sin(\Omega_0 t)\,\sigma_{\phi+\pi/2},
    \\
    u_2^\dagger(t)\sigma_\phi u_2(t)
     & =
    \sigma_\phi .
\end{align}

Introducing the Rabi-frequency error
\begin{equation}
    \Delta_i = \Omega_i-\Omega_0,
    \label{eq:Rabi_error_definition}
\end{equation}
the Hamiltonian in the second rotating frame can be written as
\begin{equation}
    H_{\mathrm{rot}}^{(2)}(t)
    =
    H_{\mathrm{err}}(t)
    +
    H_{\mathrm{co}}
    +
    H_{\mathrm{cr}}^{(2)}(t)
    +
    H_{\mathrm{int}},
    \label{eq:Hrot2_twoqubit_PMCCD_full}
\end{equation}
with
\begin{subequations}
    \label{eq:Hrot2_terms}
    \begin{align}
        H_{\mathrm{err}}(t)
         & =
        \frac{\delta_1}{2}
        \left[
            \cos(\Omega_0 t)\,\sigma_z
            +
            \sin(\Omega_0 t)\,\sigma_{\phi+\pi/2}
            \right]\otimes I
        \nonumber            \\
         & \quad
        +
        \frac{\delta_2}{2}
        I \otimes
        \left[
            \cos(\Omega_0 t)\,\sigma_z
            +
            \sin(\Omega_0 t)\,\sigma_{\phi+\pi/2}
            \right]
        \nonumber            \\
         & \quad
        +
        \frac{\Delta_1}{2}\,\sigma_\phi \otimes I
        +
        \frac{\Delta_2}{2}\,I \otimes \sigma_\phi ,
        \label{eq:Herr}
        \\
        H_{\mathrm{co}}
         & =
        \frac{\epsilon}{2}
        \left[
            \cos\theta_{\mathrm{m}}\,\sigma_z
            +
            \sin\theta_{\mathrm{m}}\,\sigma_{\phi+\pi/2}
            \right]\otimes I
        \nonumber            \\
         & \quad
        +
        \frac{\epsilon}{2}
        I \otimes
        \left[
            \cos\theta_{\mathrm{m}}\,\sigma_z
            +
            \sin\theta_{\mathrm{m}}\,\sigma_{\phi+\pi/2}
            \right],
        \label{eq:Hco}
        \\
        H_{\mathrm{cr}}^{(2)}(t)
         & =
        \frac{\epsilon}{2}
        \bigl[
            \cos(2\Omega_0 t-\theta_{\mathrm{m}})\,\sigma_z
        \nonumber            \\
         & \qquad\quad
            +
            \sin(2\Omega_0 t-\theta_{\mathrm{m}})\,\sigma_{\phi+\pi/2}
            \bigr]\otimes I
        \nonumber            \\
         & \quad
        +
        \frac{\epsilon}{2}
        I \otimes
        \bigl[
            \cos(2\Omega_0 t-\theta_{\mathrm{m}})\,\sigma_z
        \nonumber            \\
         & \qquad\qquad\quad
            +
            \sin(2\Omega_0 t-\theta_{\mathrm{m}})\,\sigma_{\phi+\pi/2}
            \bigr],
        \label{eq:Hcr}
        \\
        H_{\mathrm{int}}
         & =
        \frac{J}{4}
        \sum_{\alpha=x,y,z}
        \sigma_\alpha \otimes \sigma_\alpha .
        \label{eq:Hint}
    \end{align}
\end{subequations}
Here, $H_{\mathrm{err}}(t)$ collects the control errors: the detuning
terms, which oscillate at $\Omega_0$ in this frame, and the static
Rabi-frequency-error terms, where $\Delta_i$ describes the deviation of
the Rabi frequency of qubit $i$ from the nominal value $\Omega_0$.
Thus, $\Delta_i$ represents an individual Rabi-frequency error, whereas
the difference $\Delta_1-\Delta_2$ characterizes qubit-to-qubit
Rabi-frequency mismatch.
$H_{\mathrm{co}}$ is the static co-rotating component of the
phase-modulated CCD field; it defines the second dressed-state
quantization axis and opens an additional protection energy gap, which forms the
basis of PMCCD protection against Rabi-frequency errors.
$H_{\mathrm{cr}}^{(2)}(t)$ is the counter-rotating component of the CCD
field, oscillating at $2\Omega_0$, and $H_{\mathrm{int}}$ is the
isotropic exchange interaction.

For later discussion, we decompose the detuning and Rabi-frequency errors into common-mode and differential components,
\begin{align}
    \delta_{\mathrm{c}}
     & =
    \frac{\delta_1+\delta_2}{2},
     &
    \delta_{\mathrm{d}}
     & =
    \frac{\delta_1-\delta_2}{2},
    \label{eq:common_differential_detuning}
    \\
    \Delta_{\mathrm{c}}
     & =
    \frac{\Delta_1+\Delta_2}{2},
     &
    \Delta_{\mathrm{d}}
     & =
    \frac{\Delta_1-\Delta_2}{2}.
    \label{eq:common_differential_errors}
\end{align}
The common-mode components describe collective shifts experienced by the two qubits, whereas the differential components quantify qubit-to-qubit mismatch.
As shown below, this distinction becomes particularly important for exchange-based two-qubit operations because common-mode and differential errors affect the relative two-spin dynamics differently.

The numerical simulations presented below are performed using the full
second-rotating-frame Hamiltonian in
Eq.~\eqref{eq:Hrot2_twoqubit_PMCCD_full}, including the
counter-rotating contribution $H_{\mathrm{cr}}^{(2)}(t)$.

Finally, the isotropic Heisenberg exchange interaction retains exactly the same form under both rotating-frame transformations.
Both $U_1(t)$ and $U_2(t)$ apply identical single-qubit rotations to the two spins, and the scalar product of the spin operators is invariant under a common rotation.
Therefore, $H_{\mathrm{int}}$ has the same form in the laboratory, first rotating, and second rotating frames.
This invariance is important for the present protocol: the additional PMCCD protection energy gap can be introduced without modifying the form of the ideal isotropic exchange interaction.
It relies on the secondary field being common to the two qubits, which we assume throughout the exchange operations considered below; qubit-selective single-qubit control, in which the secondary field of one qubit is modified locally while the exchange coupling is off, is discussed in Sec.~IV and Appendix~D.
The following section examines numerically whether this additional protection remains effective during SWAP and $\sqrt{\mathrm{SWAP}}$ operations and identifies the error channels for which the protection becomes incomplete.


\section{Simulation results}
\label{sec:simulation}

\subsection{Protected SWAP dynamics}

\begin{figure*}[hbt]
    \includegraphics[width=160mm]{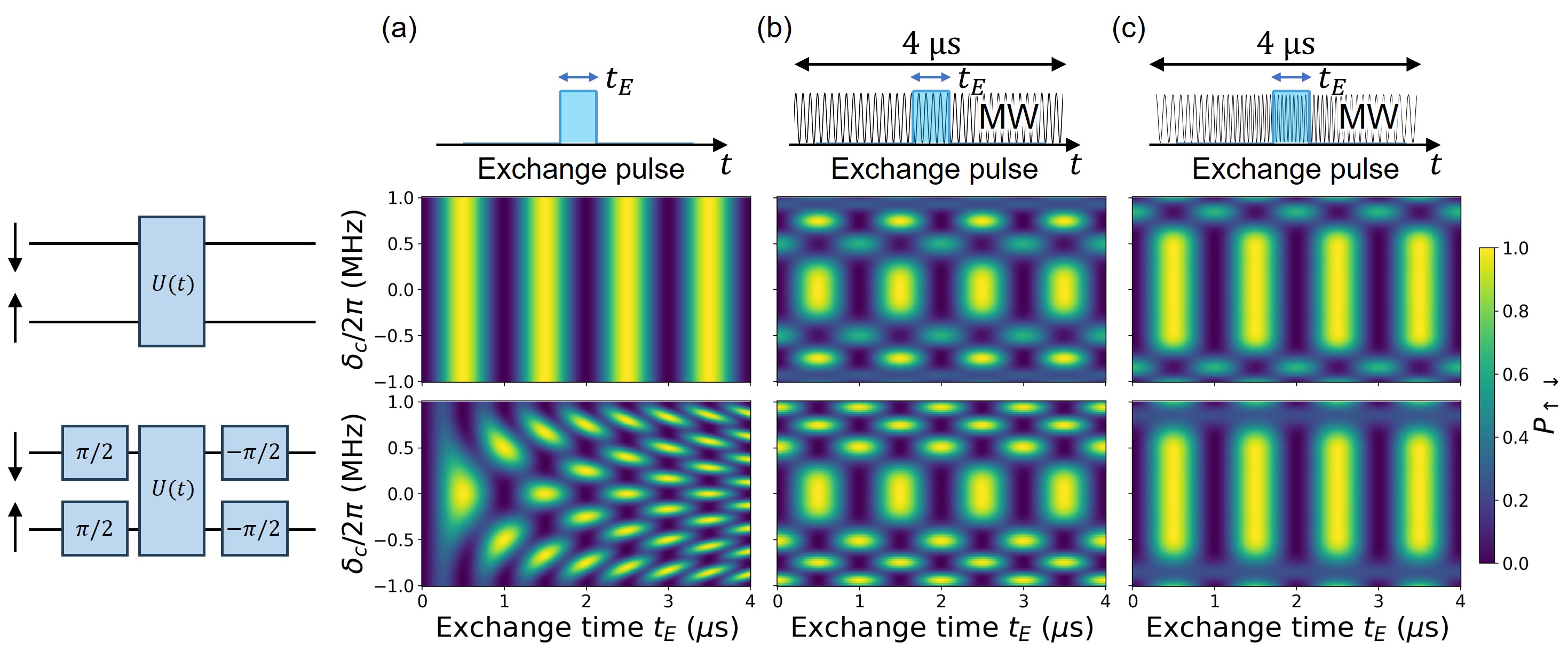}
    \caption{
    {\bf Detuning dependence of SWAP oscillations.}
    Simulated SWAP oscillations for
    (a) bare spin qubits,
    (b) conventionally dressed qubits, and
    (c) PMCCD-protected double-dressed qubits.
    The nominal Rabi angular frequency, modulation strength, and exchange coupling are set to
    $\Omega_0/2\pi = 1~\mathrm{MHz}$,
    $\epsilon=\Omega_0/4$, and
    $J/2\pi = 1~\mathrm{MHz}$, respectively.
    A common detuning is applied to the two qubits,
    $\delta_1=\delta_2$.
    The color scale shows the population $P_{\uparrow\downarrow}$ of the state $|{\uparrow\downarrow}\rangle$ for the initial state $|{\downarrow\uparrow}\rangle$ as a function of the exchange time $t_{\mathrm{E}}$.
    The upper panels show the SWAP oscillations for spin-basis states, while the lower panels show those for superposition states.
    For the lower panels, the qubits are prepared in and read out from superposition states by a collective $\pi/2$ rotation applied before and after the exchange window.
    }
    \label{swap_detuning}
\end{figure*}

We first examine the time-domain SWAP dynamics to illustrate how continuous-drive protection operates during exchange evolution.
Figure~\ref{swap_detuning} compares the detuning dependence of the SWAP oscillations for bare spin qubits, conventionally dressed qubits, and PMCCD-protected double-dressed qubits.

For the conventionally dressed and PMCCD-protected cases, the MW drive is continuously applied while the exchange interaction is turned on; in all simulations, the MW and modulation phases are set to $\theta_{\mathrm{m}}=\phi=\pi/2$.
The upper panels of Fig.~\ref{swap_detuning} show the evolution of spin-basis states, whereas the lower panels show the corresponding evolution of superposition states.
This comparison is useful because a common detuning affects not only population transfer but also the phase accumulated during the exchange operation.

For bare qubits, the population exchange between the spin-basis states is insensitive to a common detuning, which commutes with the isotropic exchange interaction.
However, superposition states acquire an additional collective phase during the exchange evolution, resulting in a deviation from the ideal SWAP operation when the full unitary evolution is considered.
Conventionally dressed qubits suppress this quasistatic phase accumulation through the dressed-state energy splitting.
PMCCD introduces an additional protection energy gap in the second rotating frame and further reduces the sensitivity of the SWAP dynamics to common detuning.
These results illustrate that continuous-drive protection can remain active while the exchange interaction generates the two-qubit operation.

\begin{figure}[hbt]
    \includegraphics[width=\columnwidth]{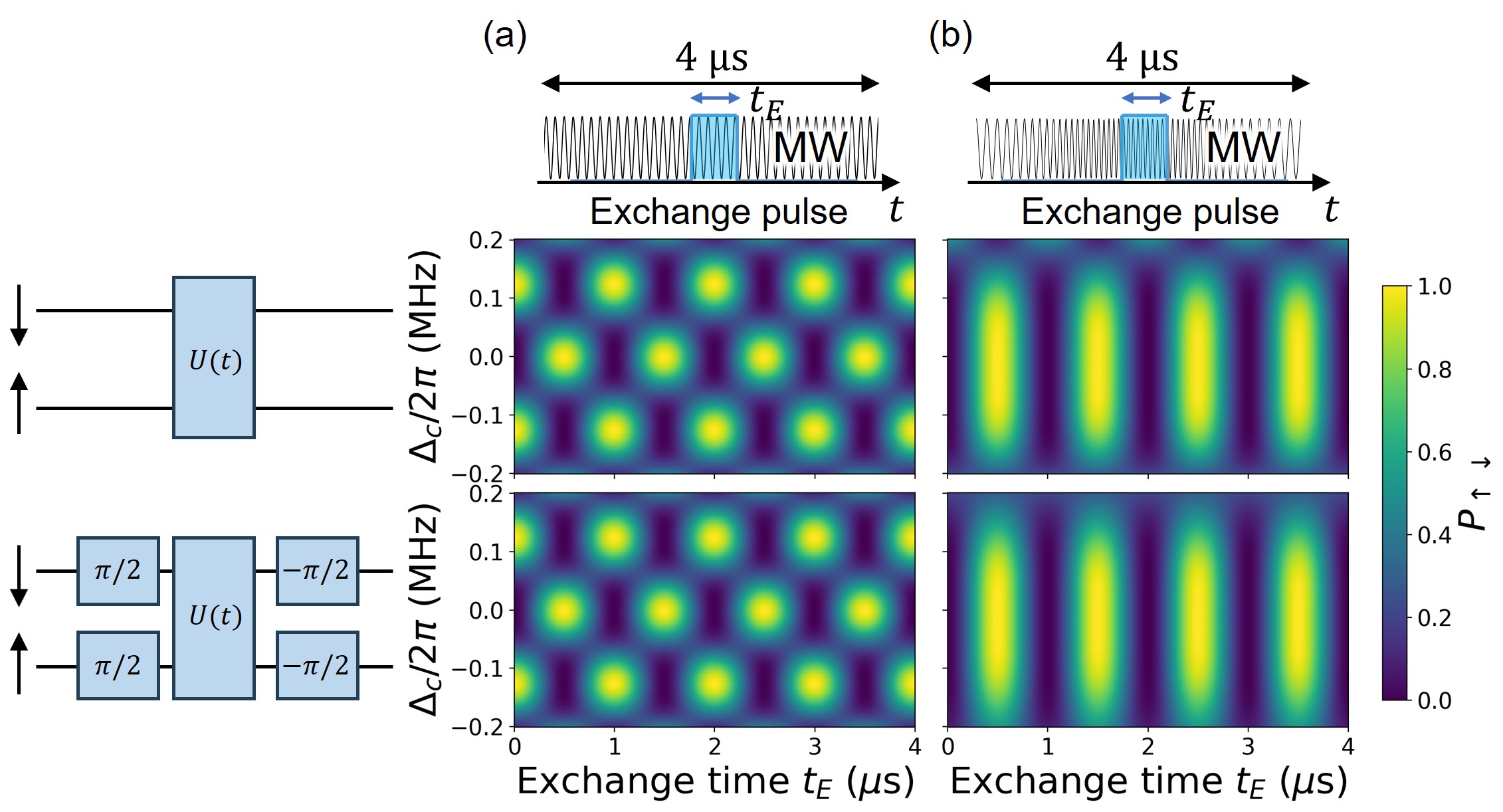}
    \caption{
    {\bf Rabi-frequency-error dependence of SWAP oscillations.}
    Simulated SWAP oscillations as a function of Rabi-frequency error for
    (a) conventionally dressed qubits and
    (b) PMCCD-protected double-dressed qubits.
    The nominal Rabi angular frequency, modulation strength, and exchange coupling are set to
    $\Omega_0/2\pi = 1~\mathrm{MHz}$,
    $\epsilon=\Omega_0/4$, and
    $J/2\pi = 1~\mathrm{MHz}$, respectively.
    A common Rabi-frequency error is applied to the two qubits,
    $\Delta_1=\Delta_2$.
    The color scale shows the population $P_{\uparrow\downarrow}$ for the initial state $|{\downarrow\uparrow}\rangle$ as a function of the exchange time $t_{\mathrm{E}}$.
    The upper panels show the SWAP oscillations for spin-basis states, while the lower panels show those for superposition states prepared and read out as in Fig.~\ref{swap_detuning}.
    }
    \label{fig:swap_rabi_error}
\end{figure}

We next examine the response to Rabi-frequency errors.
Because no MW drive is applied during the bare exchange operation, Rabi-frequency errors are not defined for the bare-qubit case.
We therefore compare conventionally dressed and PMCCD-protected double-dressed qubits in Fig.~\ref{fig:swap_rabi_error}.
Conventionally dressed qubits are directly sensitive to fluctuations of the Rabi frequency because the dressed-state splitting is determined by the MW amplitude.
In PMCCD, the secondary driving field opens an additional protection energy gap in the second rotating frame and suppresses the influence of deviations of the primary Rabi frequency from its nominal value.
As a result, the SWAP dynamics remains closer to the ideal evolution over a broader range of Rabi-frequency errors.
The time-domain results in Figs.~\ref{swap_detuning} and \ref{fig:swap_rabi_error} therefore provide a direct illustration of how the additional CCD protection can operate simultaneously with exchange-based two-qubit dynamics.

\subsection{Gate-fidelity analysis}

We next quantitatively evaluate the robustness of the exchange-based two-qubit operation.
Although the full SWAP and $\sqrt{\mathrm{SWAP}}$ operations exhibit qualitatively similar error dependence in the parameter regime considered here, we focus on the $\sqrt{\mathrm{SWAP}}$ operation because it constitutes an entangling two-qubit gate.
We quantify the gate performance using the average gate fidelity.
For the dressed and PMCCD cases, the propagator $U_{\mathrm{sim}}$ is computed
over a control window of duration $T=4~\mu\mathrm{s}$, consisting of the
exchange pulse of duration $t_{\mathrm{E}}=0.25~\mu\mathrm{s}$ for
$J/2\pi=1~\mathrm{MHz}$, followed by free evolution under the continuous
drive with the exchange coupling off; detuning and Rabi-frequency errors
act throughout the window. The window corresponds to four Rabi periods
and one period of the secondary field, $\epsilon T=2\pi$, so that the
deterministic single-qubit evolution generated by the drive returns to
the identity. The reference unitary $U_{\mathrm{ref}}$ is the propagator of
the same sequence in the absence of errors; it coincides with the ideal
$\sqrt{\mathrm{SWAP}}$ operation to within $1-F_{\mathrm{avg}}\approx
    6\times10^{-5}$, the residual being a deterministic collective rotation
caused by the counter-rotating term. For bare qubits, which are not
driven, $U_{\mathrm{sim}}$ and $U_{\mathrm{ref}}$ are evaluated over the exchange
pulse alone, so that the bare infidelities exclude errors accumulated
during idling. The simulated unitary is compared with the reference
according to~\cite{nielsen2002simple}
\begin{equation}
    F_{\mathrm{avg}}
    =
    \frac{
        \left|
        \mathrm{Tr}
        \left(
        U_{\mathrm{ref}}^{\dagger}U_{\mathrm{sim}}
        \right)
        \right|^2+d
    }{
        d(d+1)
    },
    \label{eq:Favg}
\end{equation}
where $d=4$ for the present two-qubit system.

\begin{figure}[hbt]
    \includegraphics[width=\columnwidth]{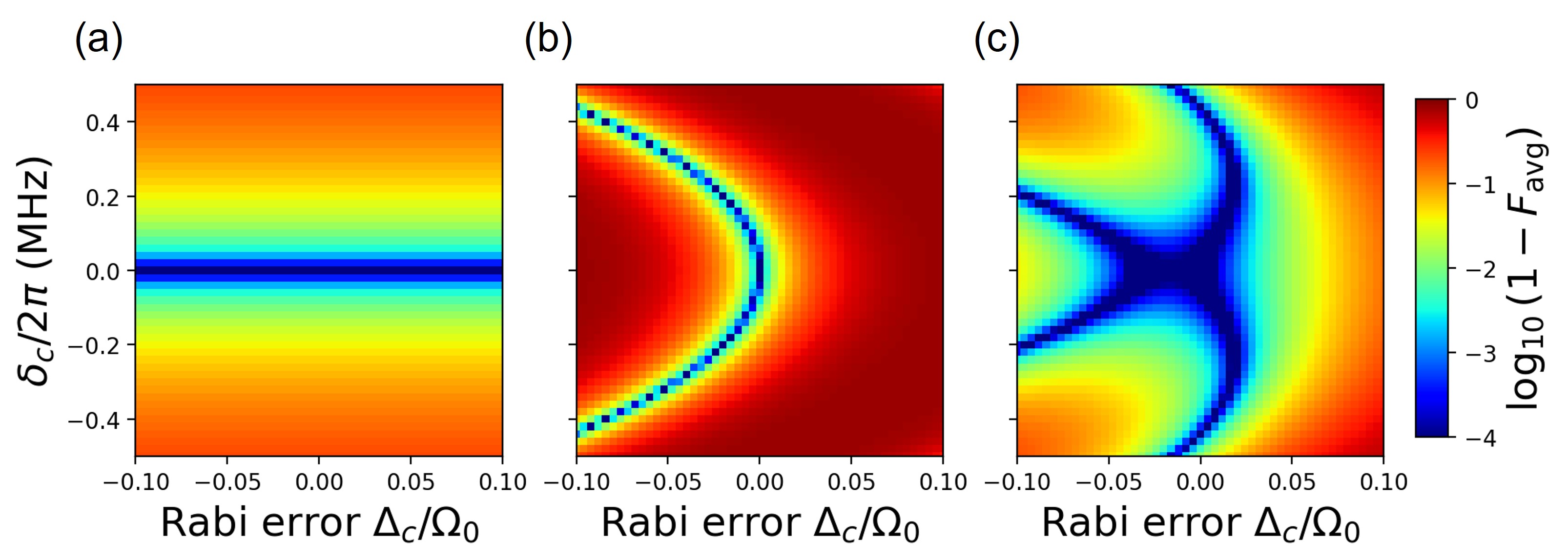}
    \caption{
    {\bf Average gate infidelity of the $\sqrt{\mathrm{SWAP}}$ operation under common-mode quasistatic errors.}
    Simulated average gate infidelity, $1-F_{\mathrm{avg}}$, as a function of common detuning and common Rabi-frequency errors for
    (a) bare spin qubits,
    (b) conventionally dressed qubits, and
    (c) PMCCD-protected double-dressed qubits.
    The nominal Rabi angular frequency, modulation strength, and exchange coupling are set to
    $\Omega_0/2\pi = 1~\mathrm{MHz}$,
    $\epsilon=\Omega_0/4$, and
    $J/2\pi = 1~\mathrm{MHz}$, respectively.
    The exchange-pulse duration is
    $t_{\mathrm{E}}=0.25~\mu\mathrm{s}$.
    The errors are applied commonly to the two qubits,
    $\delta_1=\delta_2$ and $\Delta_1=\Delta_2$.
    For the bare-qubit case, the gate fidelity is independent of $\Delta_i$ because no MW drive is applied.
    The infidelity is evaluated relative to the error-free reference unitary of each protocol.
    }
    \label{fig:fidelity_rootswap}
\end{figure}

Figure~\ref{fig:fidelity_rootswap} shows the average gate infidelity,
$1-F_{\mathrm{avg}}$, of the $\sqrt{\mathrm{SWAP}}$ operation as a function of common-mode detuning and Rabi-frequency errors.

For bare qubits, common detuning produces an uncompensated collective phase evolution in addition to the desired exchange operation and therefore increases the full gate infidelity.
Conventional dressing reduces the sensitivity to quasistatic detuning but remains susceptible to Rabi-frequency errors because fluctuations of the MW amplitude directly modify the dressed-state splitting.
PMCCD broadens the low-infidelity region along both error directions because the secondary driving field provides an additional protection energy gap against residual fluctuations of the dressed-qubit splitting.

Importantly, this comparison shows that the additional protection introduced by CCD remains effective while the exchange interaction is simultaneously generating an entangling operation.
Thus, the improvement from conventional dressing to PMCCD is not restricted to isolated single-qubit dynamics but extends to exchange-based two-qubit gates.
A comparison with amplitude- and circularly modulated CCD is given in Appendix~B.

\subsection{Detuning-error symmetry and spatial noise correlations}

The preceding analysis considered errors that are applied identically
to the two qubits. In semiconductor spin-qubit devices, qubit-frequency
fluctuations can exhibit spatial correlations between neighboring
qubits~\cite{yoneda2023noise}. For exchange-based two-qubit gates,
such correlations are particularly important because they determine
the relative contributions of common-mode and differential detuning.
A finite differential component,
$\delta_{\mathrm{d}}=(\delta_1-\delta_2)/2$, changes the relative energy of the
two exchange-coupled spins.
We therefore evaluate the $\sqrt{\mathrm{SWAP}}$ gate infidelity as a function of the independent detunings $\delta_1$ and $\delta_2$.

\begin{figure}[hbt]
    \includegraphics[width=\columnwidth]{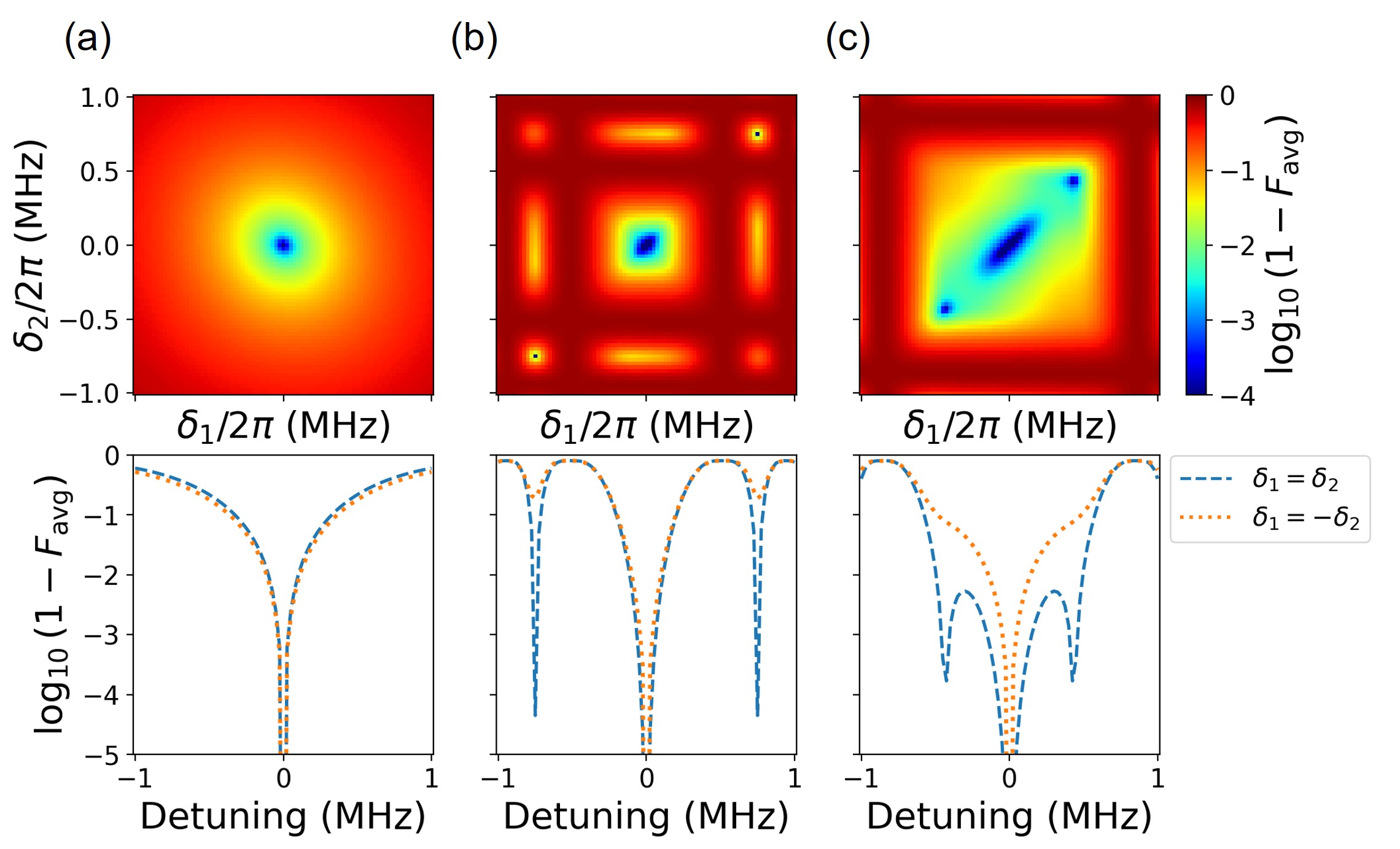}
    \caption{
    {\bf Detuning-error dependence of the $\sqrt{\mathrm{SWAP}}$ gate infidelity.}
    Simulated average gate infidelity, $1-F_{\mathrm{avg}}$, of the $\sqrt{\mathrm{SWAP}}$ operation as a function of independent detuning errors $\delta_1$ and $\delta_2$ for
    (a) bare spin qubits,
    (b) conventionally dressed qubits, and
    (c) PMCCD-protected double-dressed qubits.
    The nominal Rabi angular frequency, modulation strength, and exchange coupling are set to
    $\Omega_0/2\pi = 1~\mathrm{MHz}$,
    $\epsilon=\Omega_0/4$, and
    $J/2\pi = 1~\mathrm{MHz}$, respectively.
    The exchange-pulse duration is
    $t_{\mathrm{E}}=0.25~\mu\mathrm{s}$,
    and the Rabi-frequency errors are set to zero.
    The lower panels show one-dimensional cuts along the common-mode direction, $\delta_1=\delta_2$, and the differential direction, $\delta_1=-\delta_2$.
    }
    \label{fig:rootswap_detuning_2d}
\end{figure}

Figure~\ref{fig:rootswap_detuning_2d} reveals a pronounced difference between the common-mode and differential directions.
Along the diagonal,
$\delta_1=\delta_2$, corresponding to $\delta_{\mathrm{d}}=0$, conventionally dressed qubits exhibit a suppressed detuning sensitivity, and PMCCD further enlarges the low-infidelity region.
In contrast, along the anti-diagonal,
$\delta_1=-\delta_2$, the differential component is maximal for a given detuning amplitude, and the protection provided by continuous driving is substantially less effective.

This contrast reflects an important distinction between the two error symmetries.
Common-mode detuning predominantly generates collective single-qubit evolution, whereas differential detuning produces a relative energy mismatch between the two spins and can directly modify the exchange-based dynamics.
Consequently, the protection obtained by continuous driving depends strongly on the spatial symmetry of the noise.
The residual sensitivity to differential detuning therefore represents an important limitation of PMCCD-protected exchange gates.

Further physical insight can be obtained from a perturbative analysis of the effective exchange interaction.
As shown in Appendix~A, within a leading off-resonant high-frequency approximation, differential detuning produces an anisotropic correction to the exchange interaction, whereas the corresponding correction to the two-qubit interaction vanishes exactly for purely common-mode detuning.
This result provides a possible interaction-level mechanism contributing to the qualitatively different responses to common and differential detuning.

To connect the deterministic error maps in Fig.~\ref{fig:rootswap_detuning_2d} to a quasistatic noise model, we next consider partially correlated detuning fluctuations between the two qubits.
We model $\delta_1$ and $\delta_2$ as zero-mean Gaussian random variables with equal standard deviation $\sigma$ and correlation coefficient $\rho$, such that
\begin{equation}
    \left\langle \delta_i^2 \right\rangle=\sigma^2,
    \qquad
    \left\langle \delta_1\delta_2 \right\rangle=\rho\sigma^2.
    \label{eq:detuning_correlation}
\end{equation}
In terms of the common-mode and differential components introduced in Eq.~\eqref{eq:common_differential_detuning}, their variances are
\begin{equation}
    \left\langle \delta_{\mathrm{c}}^2 \right\rangle
    =
    \frac{\sigma^2}{2}(1+\rho),
    \qquad
    \left\langle \delta_{\mathrm{d}}^2 \right\rangle
    =
    \frac{\sigma^2}{2}(1-\rho).
    \label{eq:detuning_variances}
\end{equation}
Thus, $\rho=1$ corresponds to fully correlated fluctuations
($\delta_1=\delta_2$), whereas $\rho=-1$ corresponds to fully anticorrelated fluctuations
($\delta_1=-\delta_2$).
For $\rho=0$, the fluctuations of the two qubits are uncorrelated and the common-mode and differential components have equal variance.

We define the noise-averaged gate fidelity as
\begin{equation}
    \overline{F}_{\mathrm{avg}}(\sigma,\rho)
    =
    \left\langle
    F_{\mathrm{avg}}(\delta_1,\delta_2)
    \right\rangle_{\sigma,\rho},
    \label{eq:noise_averaged_fidelity}
\end{equation}
where the average is taken over the bivariate Gaussian distribution
\begin{equation}
    P(\delta_1,\delta_2)
    =
    \frac{1}{2\pi\sigma^2\sqrt{1-\rho^2}}
    \exp\!\left[
        -\frac{\delta_1^2-2\rho\,\delta_1\delta_2+\delta_2^2}
        {2\sigma^2(1-\rho^2)}
        \right]
    \label{eq:gaussian_joint}
\end{equation}
with the moments of Eq.~\eqref{eq:detuning_correlation}.

\begin{figure}[hbt]
    \includegraphics[width=\columnwidth]{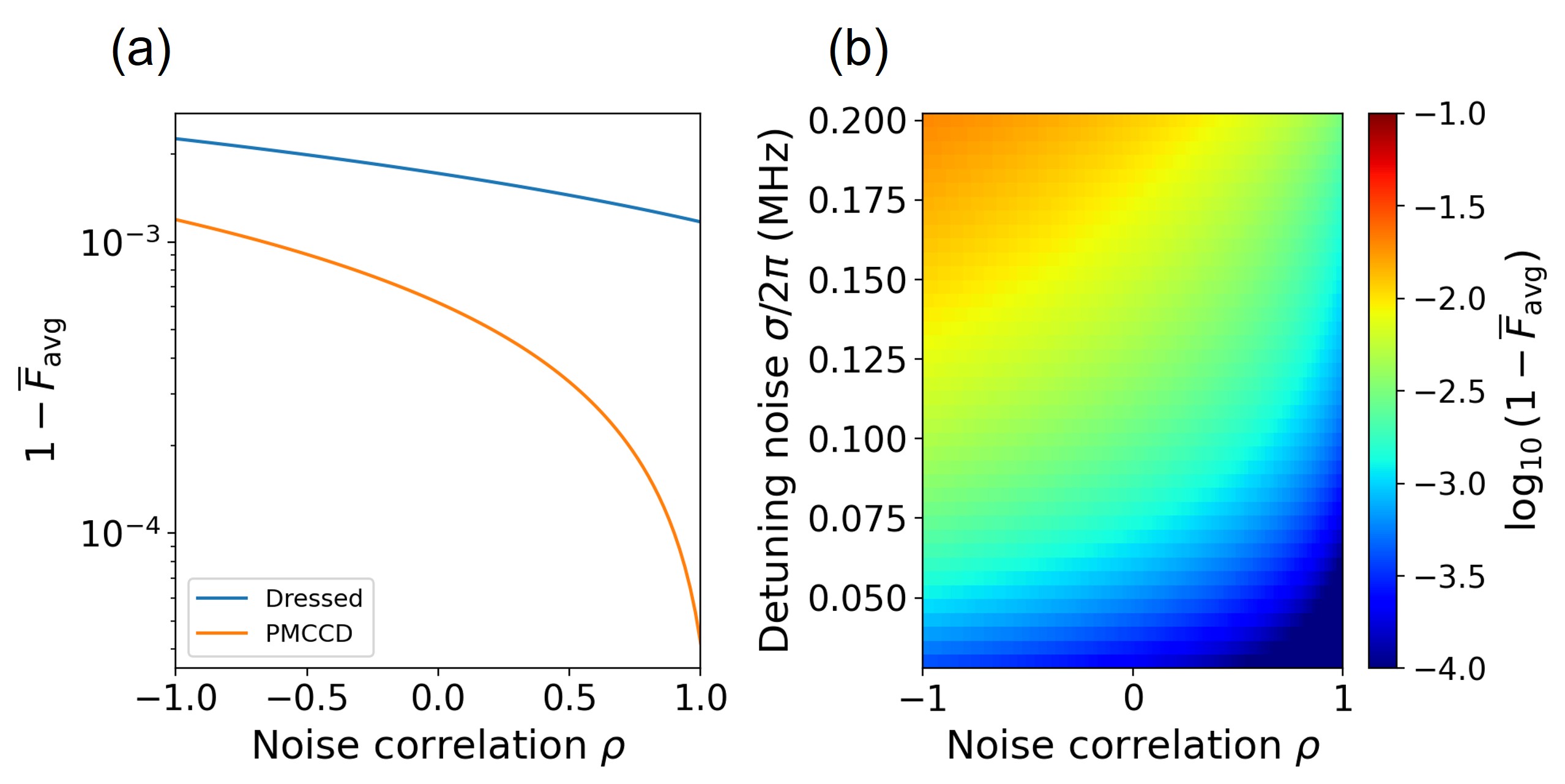}
    \caption{
    {\bf Effect of spatial correlations in quasistatic detuning noise on the $\sqrt{\mathrm{SWAP}}$ gate fidelity.}
    The detuning fluctuations $\delta_1$ and $\delta_2$ are modeled as
    zero-mean Gaussian variables with equal standard deviation $\sigma$
    and correlation coefficient $\rho$.
    (a) Noise-averaged gate infidelity $1-\overline{F}_{\mathrm{avg}}$ versus
    $\rho$ for conventionally dressed and PMCCD-protected double-dressed qubits at
    $\sigma/2\pi=0.05~\mathrm{MHz}$.
    (b) Noise-averaged infidelity of the PMCCD-protected gate as a function
    of $\rho$ and $\sigma$, with the color scale showing
    $\log_{10}(1-\overline{F}_{\mathrm{avg}})$.
    Positive correlations reduce the differential detuning component and
    therefore enhance the effectiveness of PMCCD protection.
    The averages are evaluated by Gauss--Hermite quadrature applied to a
    cubic-spline interpolation of the fidelity landscape in
    Fig.~\ref{fig:rootswap_detuning_2d}.
    }
    \label{fig:NoiseCo}
\end{figure}

Figure~\ref{fig:NoiseCo}(a) shows the resulting noise-averaged gate infidelity,
$1-\overline{F}_{\mathrm{avg}}$, as a function of $\rho$ for
$\sigma/2\pi=0.05~\mathrm{MHz}$.
For conventionally dressed qubits, the infidelity decreases only moderately as the noise becomes more positively correlated.
In contrast, the PMCCD-protected gate exhibits a much stronger dependence on $\rho$, with the infidelity decreasing by more than an order of magnitude as the noise changes from strongly anticorrelated to strongly correlated.
This behavior follows directly from the error asymmetry identified in Fig.~\ref{fig:rootswap_detuning_2d}: increasing $\rho$ reduces the variance of the differential component while increasing that of the common-mode component.
Because PMCCD efficiently suppresses common-mode detuning but remains comparatively sensitive to differential detuning, positive spatial correlations substantially enhance the protected gate fidelity.

Figure~\ref{fig:NoiseCo}(b) shows that this correlation dependence persists over a broad range of detuning-noise amplitudes.
At fixed $\sigma$, positive correlations systematically improve the PMCCD-protected gate fidelity, whereas anticorrelated noise produces substantially larger errors.
Consequently, the effectiveness of continuous-drive protection is determined not only by the overall detuning-noise amplitude but also by how the noise is partitioned between common-mode and differential components.
Spatial noise correlations therefore constitute an important parameter for assessing the expected performance of PMCCD-protected exchange gates.

\subsection{Rabi-frequency mismatch}

We next examine the robustness of the $\sqrt{\mathrm{SWAP}}$ gate against Rabi-frequency errors.
Even under a nominally global MW field, spatial inhomogeneity of the field or qubit-dependent coupling to the drive can lead to different effective Rabi frequencies for the two qubits,
$\Delta_1\neq\Delta_2$.
Such mismatch is particularly relevant for globally driven architectures because it cannot, in general, be eliminated by adjusting a single global MW amplitude.

\begin{figure}[hbt]
    \includegraphics[width=\columnwidth]{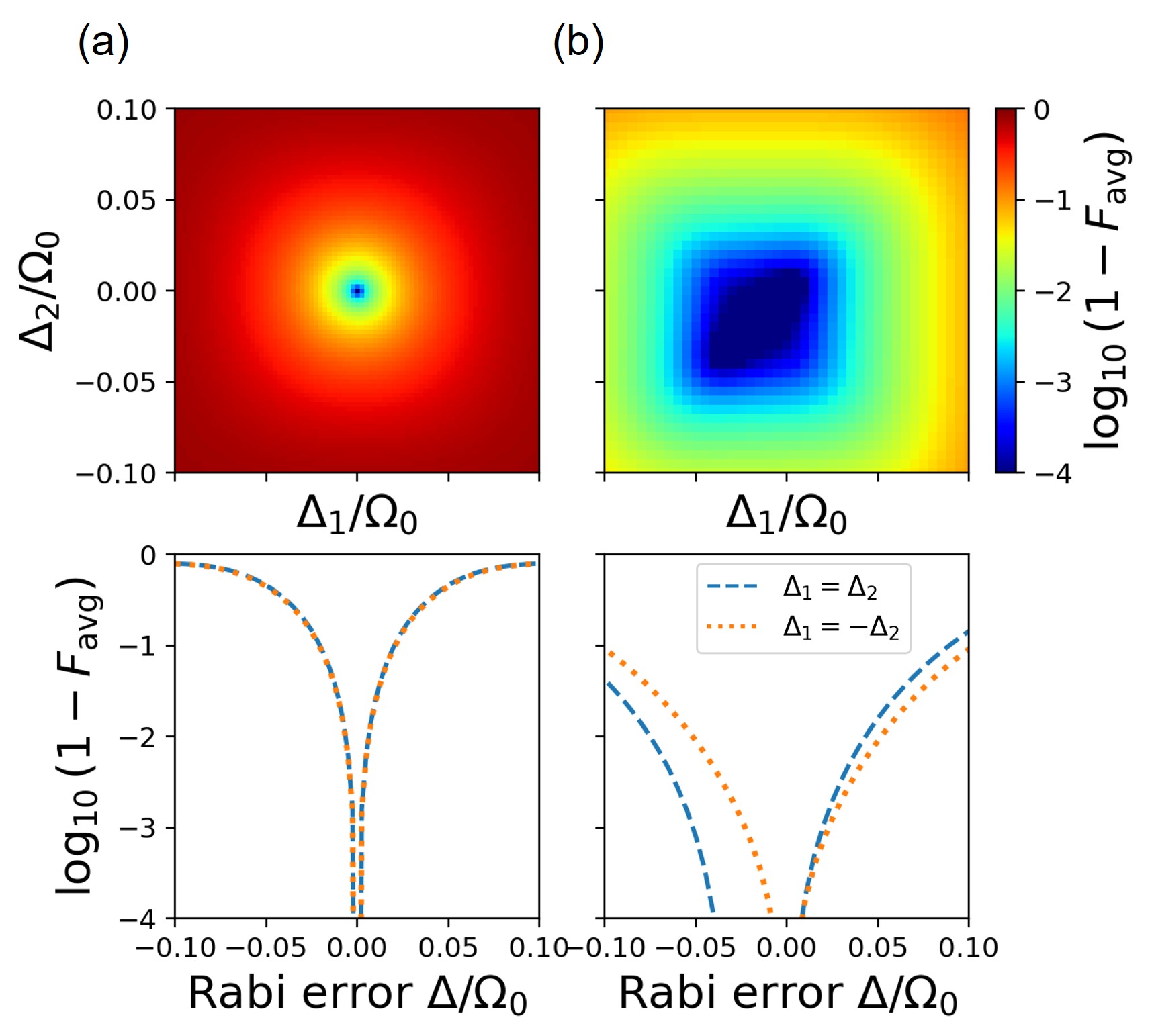}
    \caption{
    {\bf Rabi-frequency-error dependence of the $\sqrt{\mathrm{SWAP}}$ gate infidelity.}
    Simulated average gate infidelity, $1-F_{\mathrm{avg}}$, of the $\sqrt{\mathrm{SWAP}}$ operation as a function of independent Rabi-frequency errors $\Delta_1$ and $\Delta_2$ for
    (a) conventionally dressed qubits and
    (b) PMCCD-protected double-dressed qubits.
    The nominal Rabi angular frequency, modulation strength, and exchange coupling are set to
    $\Omega_0/2\pi = 1~\mathrm{MHz}$,
    $\epsilon=\Omega_0/4$, and
    $J/2\pi = 1~\mathrm{MHz}$, respectively.
    The exchange-pulse duration is
    $t_{\mathrm{E}}=0.25~\mu\mathrm{s}$,
    and the detuning errors are set to zero.
    The lower panels show cuts along $\Delta_1=\Delta_2$ and $\Delta_1=-\Delta_2$.
    }
    \label{fig:rootswap_rabi_error_2d}
\end{figure}

Figure~\ref{fig:rootswap_rabi_error_2d} shows the average gate infidelity as a function of the independent Rabi-frequency errors $\Delta_1$ and $\Delta_2$.
The diagonal direction,
$\Delta_1=\Delta_2$, corresponds to a common-mode Rabi-frequency error, whereas deviations from this diagonal contain a differential component,
$\Delta_{\mathrm{d}}\neq0$.

Conventionally dressed qubits remain sensitive to both common-mode Rabi-frequency errors and qubit-to-qubit Rabi-frequency mismatch.
In PMCCD, the additional protection energy gap suppresses deviations of the effective field in the second rotating frame and substantially enlarges the low-infidelity region.
The improvement extends away from the common-mode diagonal, demonstrating that PMCCD can provide protection not only against global MW-amplitude fluctuations but also against moderate qubit-to-qubit variations in the effective Rabi frequency.
This feature is experimentally relevant because it can relax the requirement for highly uniform MW coupling across exchange-coupled qubits.

\section{Discussion and Conclusion}

Before concluding, we discuss how the present scheme can be completed
into a universal gate set. The present analysis has considered two qubits
with matched Larmor frequencies driven by a common MW field. Since both
the global drive and the isotropic exchange interaction conserve the total
spin, the accessible operations are restricted to collective rotations and
SWAP-type gates, and completing a universal gate set requires a means of
addressing the qubits individually~\cite{loss1998quantum}. In a globally
driven architecture, qubit-specific control is available only through
the local gate voltages, which act on the qubit frequency through the
Stark shift~\cite{hansen2021pulse,hansen2024entangling}.

Qubit-selective single-qubit gates can, in principle, be implemented by
modulating the Stark shift of the addressed qubit at the Rabi frequency
$\Omega_0$. Such
a modulation produces in the rotating frame a longitudinal term of the
same form as the PMCCD secondary field
    [Eq.~\eqref{eq:Hrot1_twoqubit_PMCCD}], so that the local and global
contributions add to a single secondary field whose direction
$\theta_{\mathrm{m}}^{(i)}$, and hence the rotation axis of the double-dressed qubit,
can be set for one qubit while the others remain in the idle
configuration (Appendix~D). The secondary driving field remains present
during local control, providing a finite protection energy gap in the
double-dressed frame. The requirements are that a modulation amplitude of order $2\epsilon$
at $\Omega_0$ be applied to a local gate and that the exchange
coupling be off during the gate,
whereas the secondary fields must be kept common during exchange pulses,
as assumed in Sec.~II.

In conclusion, we have theoretically and numerically investigated
SWAP and $\sqrt{\mathrm{SWAP}}$ operations between
double-dressed spin qubits protected by phase-modulated concatenated
continuous driving (PMCCD). We find that PMCCD retains strong protection
against common-mode detuning and substantially improves robustness against
Rabi-frequency errors and qubit-to-qubit Rabi-frequency mismatch compared
with conventional continuous dressing, demonstrating that the additional
CCD protection remains effective during exchange-based entangling
operations.

The protection, however, depends strongly on the symmetry and spatial
correlation of the detuning noise. Differential detuning remains a
significant source of gate error, whereas positive spatial correlations
reduce the differential-noise component and substantially enhance PMCCD
protection. A perturbative off-resonant effective-interaction analysis
further suggests that differential detuning can dynamically renormalize the
exchange interaction and induce effective exchange anisotropy, providing
physical insight into this residual sensitivity.

These results establish the applicability of PMCCD to exchange-based
entangling gates and show that the effectiveness of continuous-drive
protection is governed not only by the magnitude of noise and control
errors but also by their symmetry, spatial correlation, and physical
origin. In contrast to qubit-frequency and Rabi-frequency errors,
exchange-coupling fluctuations directly modify the interaction generating
the gate and are not suppressed by the global continuous driving considered
here, as shown in Appendix~C. Combined with qubit-selective single-qubit
control by local Stark-shift modulation within the double-dressed frame,
which is possible in principle (Appendix~D), the protected
$\sqrt{\mathrm{SWAP}}$ gate provides the entangling element of a
universal gate set in which the qubits remain protected during idling
and entangling operations.

\appendix

\section{Effective exchange interaction under differential detuning}
\label{app:effective_exchange}

The simulations in Sec.~III show that PMCCD substantially suppresses
common-mode errors, whereas the $\sqrt{\mathrm{SWAP}}$ gate remains
more sensitive to differential detuning. To obtain physical insight
into this residual sensitivity, we examine how differential detuning
modifies the effective exchange interaction in the continuously driven
frame. The following analysis is intended as a leading-order
perturbative interpretation of the numerical results rather than as a
complete description of the gate dynamics for arbitrary parameter
regimes.

We start from the second-rotating-frame Hamiltonian of
Eq.~\eqref{eq:Hrot2_twoqubit_PMCCD_full} with the isotropic exchange
interaction $H_{\mathrm{int}}$ of Eq.~\eqref{eq:Hint}.
For clarity, we set the MW phase and CCD-modulation phase to
$\phi=\theta_{\mathrm{m}}=0$. To isolate the effect of detuning fluctuations
considered in Fig.~\ref{fig:rootswap_detuning_2d}, we further set the
Rabi-frequency errors to zero, $\Delta_1=\Delta_2=0$. For the leading
analytical treatment below, we retain the co-rotating CCD contribution
$H_{\mathrm{co}}$ and omit the counter-rotating term $H_{\mathrm{cr}}^{(2)}(t)$.
The Hamiltonian can then be written as
\begin{equation}
    H_{\mathrm{rot}}^{(2)}(t)
    =
    H_{\mathrm{s}} + V(t),
    \label{eq:App_Hrot}
\end{equation}
where the static part is
\begin{equation}
    H_{\mathrm{s}}
    =
    H_{\mathrm{co}}
    +
    H_{\mathrm{int}}
    =
    \frac{\epsilon}{2}
    \left(
    \sigma_z \otimes I
    +
    I \otimes \sigma_z
    \right)
    +
    H_{\mathrm{int}},
    \label{eq:Hstatic-SW}
\end{equation}
and the oscillating detuning contribution,
$V(t)=H_{\mathrm{err}}(t)|_{\Delta_1=\Delta_2=0}$, is
\begin{align}
    V(t)
    ={} &
    \frac{\delta_1}{2}
    \left[
        \cos(\Omega_0 t)\sigma_z
        +
        \sin(\Omega_0 t)\sigma_y
        \right]
    \otimes I
    \nonumber \\
        & +
    \frac{\delta_2}{2}
    I\otimes
    \left[
        \cos(\Omega_0 t)\sigma_z
        +
        \sin(\Omega_0 t)\sigma_y
        \right].
    \label{eq:V-detuning}
\end{align}
To obtain a leading analytical description of the effect of $V(t)$,
we consider $|\delta_i|/\Omega_0 \ll 1$ and apply a time-dependent
Schrieffer--Wolff transformation~\cite{bukov2016schrieffer},
\begin{equation}
    U_{\mathrm{SW}}(t)
    =
    \exp[-iS(t)].
    \label{eq:USW}
\end{equation}
The transformed Hamiltonian is
\begin{equation}
    \widetilde{H}(t)
    =
    U_{\mathrm{SW}}^{\dagger}(t)
    H_{\mathrm{rot}}^{(2)}(t)
    U_{\mathrm{SW}}(t)
    -
    iU_{\mathrm{SW}}^{\dagger}(t)
    \dot{U}_{\mathrm{SW}}(t).
    \label{eq:SW-transformed-H}
\end{equation}
Expanding Eq.~\eqref{eq:SW-transformed-H} to first order in the
generator gives
\begin{equation}
    \widetilde{H}(t)
    =
    H_{\mathrm{s}}
    +
    V(t)
    -
    \dot{S}(t)
    +
    i[S(t),H_{\mathrm{s}}]
    +
    \mathcal{O}(\delta_i^2).
    \label{eq:SW-first-order}
\end{equation}
Here, the generator scales as
$S(t)\sim\delta_i/\Omega_0$. In the leading high-frequency
approximation, the commutator $i[S(t),H_{\mathrm{s}}]$ is neglected when
determining the first-order generator. The oscillating perturbation is
therefore removed to this order by choosing
\begin{equation}
    \dot{S}(t)=V(t).
    \label{eq:SW-leading-condition}
\end{equation}
Since $V(t)$ has zero time average over one period, a periodic generator
with zero time average can be chosen. A convenient solution is
\begin{align}
    S(t)
    ={} &
    \frac{\delta_1}{2\Omega_0}
    \left[
        \sin(\Omega_0 t)\sigma_z
        -
        \cos(\Omega_0 t)\sigma_y
        \right]
    \otimes I
    \nonumber \\
        & +
    \frac{\delta_2}{2\Omega_0}
    I\otimes
    \left[
        \sin(\Omega_0 t)\sigma_z
        -
        \cos(\Omega_0 t)\sigma_y
        \right].
    \label{eq:SW-generator}
\end{align}
This approximation is most reliable when $\Omega_0$ is the dominant
dynamical frequency and the system is sufficiently far from resonances
involving the static energy scales. In particular, when $J$ becomes
comparable to $\Omega_0$, exchange-induced level splittings can approach
the oscillation frequency, and resonant or near-resonant contributions
neglected in the present treatment may become important. The following
result should therefore be interpreted as the leading off-resonant
correction to the two-qubit interaction.

We next expand the transformed Hamiltonian to second order in the
generator. Using $\dot{S}(t)=V(t)$, the expansion can be written as
\begin{align}
    \widetilde{H}(t)
    ={} &
    H_{\mathrm{s}}
    +
    i[S(t),H_{\mathrm{s}}]
    +
    \frac{i}{2}[S(t),V(t)]
    \nonumber \\
        & -
    \frac{1}{2}
    [S(t),[S(t),H_{\mathrm{s}}]]
    +
    \text{higher-order terms}.
    \label{eq:SW-second-order}
\end{align}
The first-order term $i[S(t),H_{\mathrm{s}}]$ has zero average over one modulation
period because $S(t)$ has zero time average. Moreover, both $S(t)$ and
$V(t)$ are sums of local single-qubit operators, so that
$[S(t),V(t)]$ generates only local single-qubit corrections.
Consequently, the leading time-averaged correction to the two-qubit
interaction is
\begin{equation}
    \Delta H_{\mathrm{int}}
    =
    -\frac{1}{2}
    \left\langle
    [S(t),[S(t),H_{\mathrm{int}}]]
    \right\rangle_t ,
    \label{eq:interaction-correction}
\end{equation}
where $\langle\cdots\rangle_t$ denotes averaging over one period,
$2\pi/\Omega_0$.

Using
\begin{equation}
    \left\langle\sin^2(\Omega_0 t)\right\rangle_t
    =
    \left\langle\cos^2(\Omega_0 t)\right\rangle_t
    =
    \frac{1}{2},
\end{equation}
together with
\begin{equation}
    \left\langle
    \sin(\Omega_0 t)\cos(\Omega_0 t)
    \right\rangle_t
    =
    0,
\end{equation}
we obtain
\begin{equation}
    \Delta H_{\mathrm{int}}
    =
    -\frac{J(\delta_1-\delta_2)^2}{16\Omega_0^2}
    \left(
    2\sigma_x\otimes\sigma_x
    +
    \sigma_y\otimes\sigma_y
    +
    \sigma_z\otimes\sigma_z
    \right).
    \label{eq:dynamic-correction}
\end{equation}

A key feature of Eq.~\eqref{eq:dynamic-correction} is that the
interaction correction depends only on the differential detuning,
$\delta_1-\delta_2=2\delta_{\mathrm{d}}$ [Eq.~\eqref{eq:common_differential_detuning}].
For a purely common-mode
fluctuation, $\delta_1=\delta_2$, the correction to the two-qubit
interaction therefore vanishes. This holds to all orders, because
$U_{\mathrm{SW}}(t)$ then applies the same rotation to both spins, under
which the isotropic exchange interaction is invariant. This result is
consistent with the
qualitatively different common-mode and differential responses observed
in Fig.~\ref{fig:rootswap_detuning_2d}.

The absence of an interaction correction for common-mode detuning does
not imply that common-mode fluctuations produce no gate error. They can
still generate collective single-qubit evolution and modify the full
two-qubit unitary operation. In contrast, differential detuning changes
the relative dynamics of the two spins and directly renormalizes the
exchange interaction within the present perturbative description.

We emphasize that Eq.~\eqref{eq:dynamic-correction} contains only the
correction to the two-qubit interaction. The same transformation
also generates local dressed-qubit corrections, which are not retained
here because our purpose is to identify the interaction-level mechanism
associated with differential detuning. Within the present leading-order
treatment of the two-qubit interaction, the correction does not explicitly
depend on the secondary protection strength $\epsilon$, indicating a
residual interaction-level contribution associated with differential
detuning.

Adding Eq.~\eqref{eq:dynamic-correction} to the isotropic exchange
interaction gives
\begin{equation}
    H_{\mathrm{int,eff}}
    =
    J_{x,\mathrm{eff}}\,
    \sigma_x\otimes\sigma_x
    +
    J_{y,\mathrm{eff}}\,
    \sigma_y\otimes\sigma_y
    +
    J_{z,\mathrm{eff}}\,
    \sigma_z\otimes\sigma_z,
    \label{eq:Hint-effective}
\end{equation}
where
\begin{align}
    J_{x,\mathrm{eff}}
     & =
    \frac{J}{4}
    -
    \frac{J(\delta_1-\delta_2)^2}{8\Omega_0^2},
    \label{eq:Jx-dynamic}
    \\
    J_{y,\mathrm{eff}}
     & =
    \frac{J}{4}
    -
    \frac{J(\delta_1-\delta_2)^2}{16\Omega_0^2},
    \label{eq:Jy-dynamic}
    \\
    J_{z,\mathrm{eff}}
     & =
    \frac{J}{4}
    -
    \frac{J(\delta_1-\delta_2)^2}{16\Omega_0^2}.
    \label{eq:Jz-dynamic}
\end{align}
Within the leading off-resonant high-frequency approximation,
Eqs.~\eqref{eq:Jx-dynamic}--\eqref{eq:Jz-dynamic} show that
differential detuning produces unequal renormalization of the exchange
couplings. In particular, for the choice $\phi=0$ used in the present
derivation, the correction to the coupling along the primary-drive
$x$ axis is twice that along the transverse $y$ and $z$ axes,
\begin{equation}
    J_{x,\mathrm{eff}}-J_{y,\mathrm{eff}}
    =
    -\frac{J(\delta_1-\delta_2)^2}{16\Omega_0^2}.
    \label{eq:exchange-anisotropy}
\end{equation}
Thus, an initially isotropic Heisenberg interaction acquires an
effective dynamical exchange anisotropy in the presence of
differential detuning.

This perturbative result provides a physical interpretation of the
residual differential-detuning sensitivity observed in the numerical
simulations. Whereas common-mode fluctuations primarily contribute
collective single-qubit evolution, differential fluctuations can modify
the interaction itself and thereby perturb the intended exchange-based
gate. The resulting dynamical anisotropy can therefore contribute to
the degradation of the $\sqrt{\mathrm{SWAP}}$ gate fidelity.

These results provide a qualitative interpretation of the
differential-detuning sensitivity observed in the simulations, rather
than a complete quantitative description of the gate error.

\section{Comparison of concatenated continuous-driving protocols}
\label{app:ccd_comparison}

We compare three implementations of concatenated continuous driving:
amplitude-modulated CCD (AMCCD), phase-modulated CCD (PMCCD), and
circularly modulated CCD (CMCCD). In AMCCD, the secondary driving field
is generated by modulation of the MW amplitude, whereas in PMCCD it is
generated by modulation of the MW phase.
CMCCD combines amplitude and
phase modulation with equal weight, producing a circularly polarized
modulation field in the first rotating frame in the absence of control
errors~\cite{kuno2026concatenated}.

Following Ref.~\cite{kuno2026concatenated}, we introduce the dimensionless parameters
$\alpha_A$ and $\alpha_P$ to specify the relative contributions of
amplitude and phase modulation, respectively, with
$\alpha_A+\alpha_P=1$. For a single qubit, the generalized
laboratory-frame Hamiltonian is written as
\begin{widetext}
    \begin{equation}
        \begin{aligned}
            H_{\mathrm{lab}}(t)
            ={} &
            \frac{\omega_{\mathrm{L}}}{2}\sigma_z
            +(\Omega_0+\Delta_\Omega)
            \Bigg[
                \cos\left(
                \omega_{0}t+\phi
                -\frac{2\alpha_P\epsilon}{\Omega_0}
                \sin(\Omega_0t-\theta_{\mathrm{m}})
                \right)
            \\
                & \qquad
                +\frac{2\alpha_A\epsilon}{\Omega_0}
                \sin(\Omega_0t-\theta_{\mathrm{m}})
                \sin\left(
                \omega_{0}t+\phi
                -\frac{2\alpha_P\epsilon}{\Omega_0}
                \sin(\Omega_0t-\theta_{\mathrm{m}})
                \right)
                \Bigg]\sigma_x .
        \end{aligned}
        \label{eq:generalized_ccd}
    \end{equation}
\end{widetext}

Here, $\omega_{\mathrm{L}}$ is the qubit Larmor angular frequency,
$\Omega_0$ is the nominal Rabi angular frequency, $\Delta_{\Omega}$
denotes the Rabi-frequency error, and $\epsilon$ and
$\theta_{\mathrm{m}}$ are the modulation strength and modulation phase,
respectively. The parameters $\alpha_A$ and $\alpha_P$ determine the
relative weights of amplitude and phase modulation. The three CCD
protocols are specified by
\begin{equation}
    (\alpha_A,\alpha_P)=
    \begin{cases}
        (1,0),     & \mathrm{AMCCD}, \\
        (0,1),     & \mathrm{PMCCD}, \\
        (1/2,1/2), & \mathrm{CMCCD}.
    \end{cases}
    \label{eq:ccd_protocols}
\end{equation}

For the two-qubit calculations, $\Delta_{\Omega}$ in
Eq.~\eqref{eq:generalized_ccd} is replaced by the corresponding
qubit-dependent Rabi-frequency error $\Delta_i$. In the second rotating
frame, the coefficients of the co-rotating and counter-rotating
components of the secondary driving field for qubit $i$ are
\begin{align}
    C_{\mathrm{co}}^{(i)}
     & =
    \alpha_P+
    \left(
    1+\frac{\Delta_i}{\Omega_0}
    \right)\alpha_A,
    \\
    C_{\mathrm{cr}}^{(i)}
     & =
    \alpha_P-
    \left(
    1+\frac{\Delta_i}{\Omega_0}
    \right)\alpha_A,
\end{align}
respectively, where the corresponding field amplitudes are
$C_{\mathrm{co}}^{(i)}\epsilon/2$ and
$C_{\mathrm{cr}}^{(i)}\epsilon/2$.
In the absence of Rabi-frequency errors, $\Delta_i=0$, all three
protocols have the same co-rotating secondary-field strength because
$\alpha_A+\alpha_P=1$. In contrast, the counter-rotating component has
opposite signs for AMCCD and PMCCD and vanishes for CMCCD.

For all three protocols, we use the same modulation
strength, $\epsilon=\Omega_0/4$, so that their error-free co-rotating
secondary protection fields have the same magnitude, consistent with
the single-qubit comparison in
Ref.~\cite{kuno2026concatenated}.

\begin{figure}[hbt]
    \includegraphics[width=\columnwidth]
    {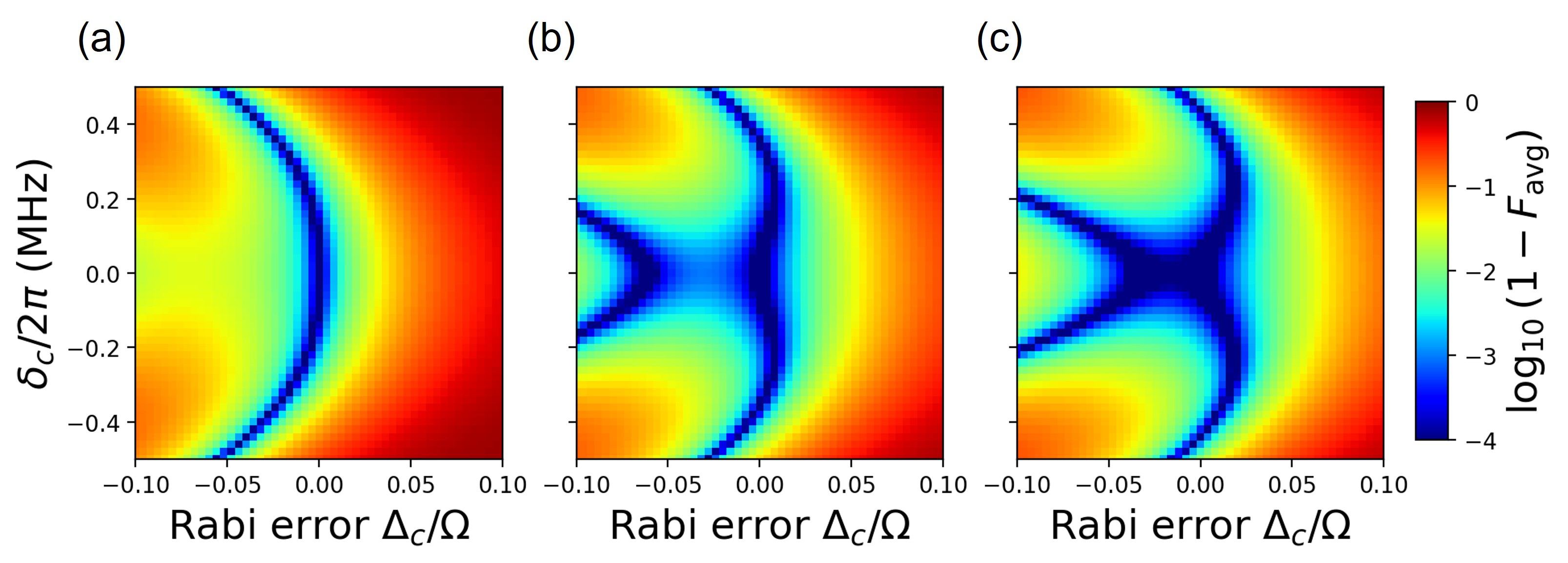}
    \caption{
    {\bf Average gate infidelity of the $\sqrt{\mathrm{SWAP}}$ operation for different CCD schemes.}
    Simulated average gate infidelity, $1-F_{\mathrm{avg}}$, of the
    $\sqrt{\mathrm{SWAP}}$ operation as a function of common detuning
    and common Rabi-frequency errors for
    (a) AMCCD,
    (b) CMCCD, and
    (c) PMCCD.
    The nominal Rabi angular frequency, modulation strength, and
    exchange coupling are set to
    $\Omega_0/2\pi=1~\mathrm{MHz}$,
    $\epsilon=\Omega_0/4$, and
    $J/2\pi=1~\mathrm{MHz}$, respectively.
    The exchange-pulse duration is
    $t_{\mathrm{E}}=0.25~\mu\mathrm{s}$, corresponding to a
    $\sqrt{\mathrm{SWAP}}$ operation.
    The errors are applied commonly to the two qubits,
    $\delta_1=\delta_2$ and $\Delta_1=\Delta_2$.
    }
    \label{fig:fidelity_am_cm_pm_ccd_sqrt_swap}
\end{figure}

As shown in Fig.~\ref{fig:fidelity_am_cm_pm_ccd_sqrt_swap}, the three
protocols exhibit qualitatively similar robustness against common-mode
detuning. Their responses to Rabi-frequency errors, however,
differ appreciably. For the parameters considered here, PMCCD exhibits
the broadest low-infidelity region. This behavior is consistent with the
different dependence of the secondary driving field on the Rabi-frequency
error. For PMCCD, $\alpha_A=0$, so that the co-rotating secondary-field
amplitude is independent of $\Delta_i$. By contrast, AMCCD and CMCCD
contain an amplitude-modulated component, and their co-rotating fields
therefore acquire an explicit dependence on $\Delta_i$. This provides
a natural explanation for the enhanced tolerance of PMCCD to common
Rabi-frequency errors in the exchange-based two-qubit operation
considered here.

\section{Sensitivity to exchange-coupling fluctuations}
\label{app:J_error}

We finally examine the effect of quasistatic fluctuations of the
exchange coupling. Such fluctuations are particularly relevant to
semiconductor spin qubits because the exchange interaction is
electrically controlled and can therefore be affected by charge
noise~\cite{reed2016reduced}. We write the exchange coupling as
\begin{equation}
    J = J_0 + \delta J,
\end{equation}
where $J_0$ is the nominal exchange coupling and $\delta J$ denotes a
quasistatic exchange-coupling error. In the following calculations, the
detuning and Rabi-frequency errors are set to zero in order to isolate
the effect of $\delta J$.

\begin{figure}[hbt]
    \includegraphics[width=\columnwidth]{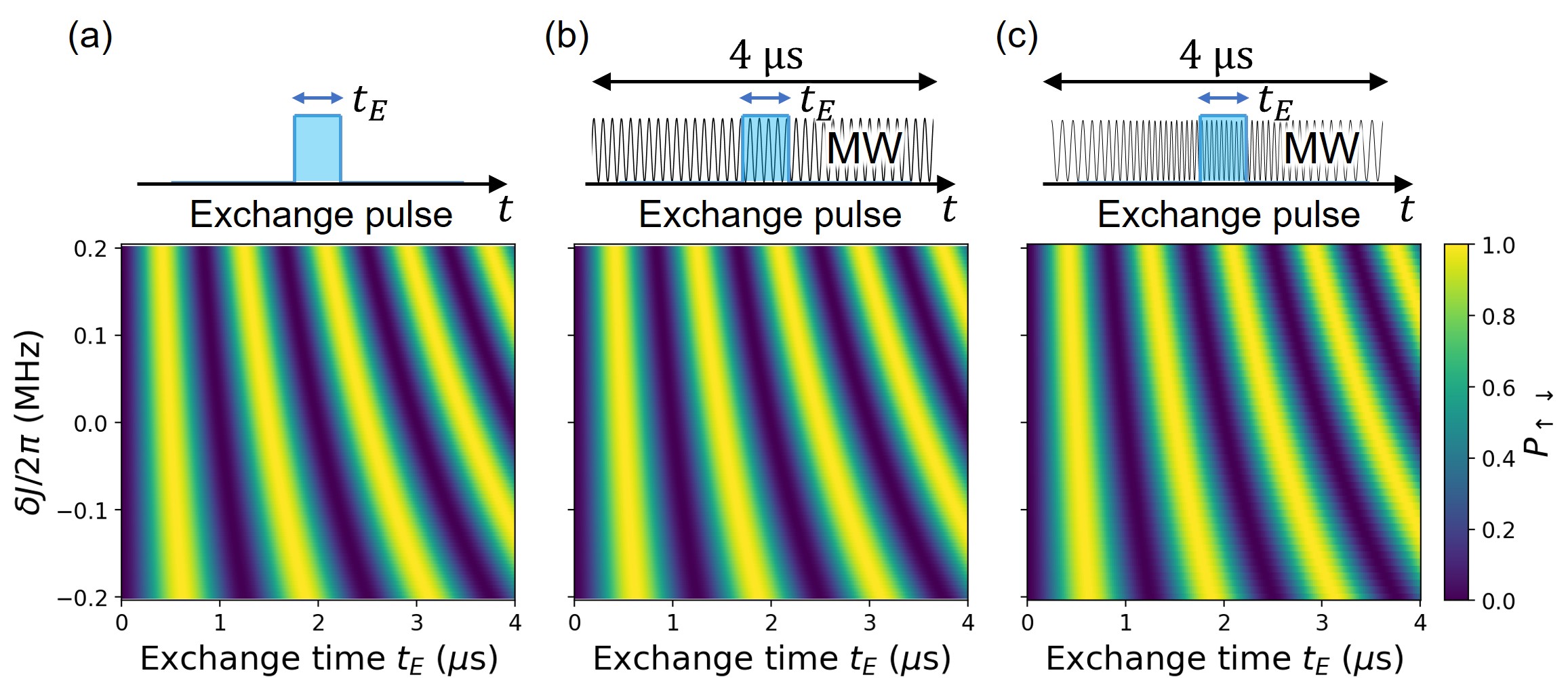}
    \caption{
    {\bf Exchange-coupling-error dependence of SWAP oscillations.}
    Simulated SWAP dynamics as a function of quasistatic exchange-coupling
    error $\delta J$ for
    (a) bare spin qubits,
    (b) conventionally dressed qubits, and
    (c) PMCCD-protected double-dressed qubits.
    The exchange coupling is written as $J=J_0+\delta J$, with
    $J_0/2\pi=1~\mathrm{MHz}$.
    For the dressed and PMCCD-protected cases,
    $\Omega_0/2\pi=1~\mathrm{MHz}$, and for PMCCD
    $\epsilon=\Omega_0/4$.
    The detuning and Rabi-frequency errors are set to zero.
    The color scale represents the population $P_{\uparrow\downarrow}$ for the initial state
    $|{\downarrow\uparrow}\rangle$ as a function of the exchange time $t_{\mathrm{E}}$.
    The nearly identical dependence in the three cases demonstrates that
    continuous driving does not suppress fluctuations of the exchange
    coupling.
    }
    \label{fig:J_error}
\end{figure}

Figure~\ref{fig:J_error} compares the SWAP dynamics for bare,
conventionally dressed, and PMCCD-protected qubits as a function of
$\delta J$. In all three cases, the exchange-coupling error changes the
oscillation frequency and therefore shifts the time required to realize
the desired exchange rotation. In contrast to detuning and
Rabi-frequency errors, however, neither conventional dressing nor PMCCD
reduces the sensitivity to $\delta J$. The three protocols exhibit
essentially the same dependence on the exchange-coupling error.

This behavior follows directly from the rotational invariance of the
isotropic Heisenberg interaction. The error associated with the exchange
fluctuation is
\begin{equation}
    H_{\delta J}
    =
    \frac{\delta J}{4}
    \sum_{\alpha=x,y,z}
    \sigma_\alpha\otimes\sigma_\alpha .
\end{equation}

As discussed in Sec.~II, the collective rotations generated by the global
continuous drive leave the isotropic exchange interaction invariant.
Consequently, $H_{\delta J}$ is also invariant under the transformations
to the dressed and double-dressed frames. The exchange-coupling error is
therefore not converted into an oscillating perturbation that can be
averaged out by continuous driving. Instead, it directly changes the
exchange rotation angle accumulated during the gate.

For a $\sqrt{\mathrm{SWAP}}$ gate with the exchange-pulse duration fixed
at its nominal value,
\begin{equation}
    t_{\mathrm{E}}=\frac{\pi}{2J_0},
\end{equation}
the resulting average gate infidelity is
\begin{equation}
    1-F_{\mathrm{avg}}
    =
    \frac{3}{5}
    \sin^2\left(
    \frac{\delta J\,t_{\mathrm{E}}}{2}
    \right)
    =
    \frac{3}{5}
    \sin^2\left(
    \frac{\pi}{4}\frac{\delta J}{J_0}
    \right).
\end{equation}

Thus, within the present model, the sensitivity to quasistatic
exchange-coupling fluctuations is identical for bare, conventionally
dressed, and PMCCD-protected exchange gates. This result highlights an
important limitation of continuous-drive protection: while dressing can
suppress errors associated with single-qubit frequency and MW-control
fluctuations, it does not suppress fluctuations of the interaction that
directly generates the two-qubit gate.

\section{Qubit-selective control by local Stark-shift modulation}
\label{app:local_addressing}

In this appendix we show that the secondary field of PMCCD can be
generated, and its direction $\theta_{\mathrm{m}}$ controlled, for an individual
qubit by modulating the qubit frequency through the electrically
controlled Stark shift while the global drive is maintained. Except in
the final paragraph, the exchange coupling is assumed to be off.

Consider qubit $i$ whose Larmor frequency is modulated as
\begin{equation}
    \omega_i(t)
    =
    \omega_i
    +
    A_i\cos\!\big(\Omega_0 t-\theta_{\mathrm{s}}^{(i)}\big),
    \label{eq:stark_modulation}
\end{equation}
where, assuming a linear Stark response,
$A_i=(\partial\omega_i/\partial V_i)\,V_{\mathrm{m},i}$ is set by the amplitude
$V_{\mathrm{m},i}$ of a sinusoidal voltage applied to a local gate.
Repeating the transformation of Sec.~II to the frame following the
phase-modulated carrier, the single-qubit part of the first-rotating-frame
Hamiltonian of qubit $i$ becomes
\begin{align}
    h_i^{(1)}(t)
     & =
    \frac{\delta_i}{2}\,\sigma_z
    +
    \frac{\Omega_i}{2}\,\sigma_\phi
    \nonumber \\
     & \quad
    +
    \Big[
        \epsilon\cos(\Omega_0 t-\theta_{\mathrm{m}})
        +
        \frac{A_i}{2}\cos\!\big(\Omega_0 t-\theta_{\mathrm{s}}^{(i)}\big)
        \Big]\sigma_z .
    \label{eq:h1_local}
\end{align}
The first term in the square brackets is the global secondary field of
Eq.~\eqref{eq:Hrot1_twoqubit_PMCCD}; the second is the local
contribution. The two terms have the same operator form because the
rotating frame depends only on the difference between the qubit and
carrier frequencies: the global phase modulation shifts the carrier
frequency by $-2\epsilon\cos(\Omega_0 t-\theta_{\mathrm{m}})$
[Eq.~\eqref{eq:Phi_CCD_twoqubit}], whereas the Stark modulation shifts
the qubit frequency by $+A_i\cos(\Omega_0 t-\theta_{\mathrm{s}}^{(i)})$. A local
modulation with $A_i=2\epsilon$ and $\theta_{\mathrm{s}}^{(i)}=\theta_{\mathrm{m}}$ therefore
reproduces the PMCCD secondary field exactly, including its
counter-rotating component, within the same rotating-wave approximation
used for the global drive. In particular, the secondary field could be
generated entirely locally, without phase modulation of the global
carrier.

Since the global and local contributions in Eq.~\eqref{eq:h1_local}
oscillate at the same frequency, they add to a single sinusoid,
\begin{equation}
    \epsilon\cos(\Omega_0 t-\theta_{\mathrm{m}})
    +
    \frac{A_i}{2}\cos\!\big(\Omega_0 t-\theta_{\mathrm{s}}^{(i)}\big)
    =
    \epsilon_i^{\mathrm{eff}}\cos\!\big(\Omega_0 t-\theta_{\mathrm{m}}^{(i)}\big),
    \label{eq:coherent_sum}
\end{equation}
with amplitude and phase given by
\begin{equation}
    \epsilon_i^{\mathrm{eff}}\,e^{i\theta_{\mathrm{m}}^{(i)}}
    =
    \epsilon\,e^{i\theta_{\mathrm{m}}}
    +
    \frac{A_i}{2}\,e^{i\theta_{\mathrm{s}}^{(i)}} .
    \label{eq:complex_sum}
\end{equation}
The amplitude and direction of the secondary field of qubit $i$ are thus
set by the local waveform, whereas qubits without local modulation retain
$(\epsilon,\theta_{\mathrm{m}})$. To rotate the secondary field of qubit $i$ from
$\theta_{\mathrm{m}}$ to $\theta_{\mathrm{m}}+\theta$ at unchanged amplitude $\epsilon$,
Eq.~\eqref{eq:complex_sum} requires
\begin{equation}
    A_i
    =
    4\epsilon\sin\frac{\theta}{2},
    \qquad
    \theta_{\mathrm{s}}^{(i)}
    =
    \theta_{\mathrm{m}}+\frac{\pi}{2}+\frac{\theta}{2}.
    \label{eq:local_waveform}
\end{equation}
For $\theta=\pi/2$ this gives $A_i=2\sqrt{2}\,\epsilon$ and
$\theta_{\mathrm{s}}^{(i)}=\theta_{\mathrm{m}}+3\pi/4$; for $\theta=\pi$, $A_i=4\epsilon$ and
$\theta_{\mathrm{s}}^{(i)}=\theta_{\mathrm{m}}+\pi$. Alternatively, adding a local component in
quadrature, $\theta_{\mathrm{s}}^{(i)}=\theta_{\mathrm{m}}+\pi/2$, without canceling the global
one tilts the axis by $\arctan(A_i/2\epsilon)$ at amplitude
$\sqrt{\epsilon^2+A_i^2/4}$, which together with $Z$ rotations is
sufficient for universal control at a reduced modulation amplitude.

Transforming to the second rotating frame as in Sec.~II, the secondary
field of qubit $i$ becomes the static field
\begin{equation}
    \frac{\epsilon_i^{\mathrm{eff}}}{2}
    \left[
        \cos\theta_{\mathrm{m}}^{(i)}\,\sigma_z
        +
        \sin\theta_{\mathrm{m}}^{(i)}\,\sigma_{\phi+\pi/2}
        \right],
    \label{eq:second_frame_local}
\end{equation}
that is, the single-qubit part of $H_{\mathrm{co}}$ in
Eq.~\eqref{eq:Hco} with $(\epsilon,\theta_{\mathrm{m}})$ replaced by
$(\epsilon_i^{\mathrm{eff}},\theta_{\mathrm{m}}^{(i)})$, plus a
counter-rotating term at $2\Omega_0$ analogous to
Eq.~\eqref{eq:Hcr}. The double-dressed qubit precesses about this axis at
angular frequency $\epsilon_i^{\mathrm{eff}}$. Taking $\theta_{\mathrm{m}}=0$
as the idle reference, setting $\theta_{\mathrm{m}}^{(i)}=\pi/2$ generates
a rotation of qubit $i$ about $\sigma_{\phi+\pi/2}$, while
$\theta_{\mathrm{m}}^{(i)}=\pi$ produces a $Z$ rotation relative to the
other qubits. These noncommuting rotations enable arbitrary single-qubit
control. The secondary protection energy gap remains present during local
control.

During exchange pulses, the secondary fields must be identical for both
qubits, as assumed in Sec.~II.

\section*{Acknowledgments}
This work was supported by JST Moonshot R\&D Grant No.~JPMJMS256H.


\clearpage

\bibliographystyle{apsrev4-2}
\bibliography{refs}

\end{document}